\documentclass{article}[10pt]

\usepackage{jucs2e}
\usepackage{url}
\usepackage{graphicx} 
\usepackage{setspace}
\usepackage{multirow}
\begin{document}

\title{Hybrid Content   Dynamic  Recommendation  System 
Based in Adapted Tags and Applied to Digital Library
}

\author{{\bfseries Thiago Bellotti Furtado\\
   (Federal University of Lavras, Lavras, Brazil\\
   thiagofurtado@ufla.br)
   \and
   {\bfseries Ahmed Esmin}\\
   (Federal University of Lavras, Lavras, Brazil  \\
   ahmed@ufla.br)\\
}}
\maketitle

\begin{abstract}
The technological evolution of the library in the academic environment brought a lot of information and documents that are available to access, but these systems do not always have mechanisms to search in an integrated way the relevant information for the user. To alleviate this problem, we propose a recommendation system that generates the user profile through tags that are reshaped over time. To trace the user profile the system uses information from your lending history stored in the library database and it collects their opinions (feedback) through a list of recommendations. These data are integrated with the document base of institutional repository. Thus, the recommendation system assists users in identifying relevant items and makes suggestions for content in an integrated environment that contains institutional repository documents and the university library database. The proposed recommendation system uses a hybrid approach being applied in an academic environment with the participation of the users.
\end{abstract}


\section{Introduction}

Technological changes and recent concepts of information resources management have caused a break in the paradigm of traditional models of libraries. The concept of digital libraries presents an alternative to enlarge the search conditions, availability and the information retrieval in a globalized way, mainly relying on access to web networks. The advance in this type of technology has provided people and organizations access to a large amount of documents and information. However, in most cases this potential is not correctly exploited. This is because there are some complicating factors, such as lack of integration among systems that do not have mechanisms to facilitate the search for new information \cite{tejeda-Lorente2014}.

For storing large volumes of documents and information that are available to the entire academic community, the universities’ digital library systems, also known as Institutional Repositories (IR) have become an important tool for the democratization of scientific production, supporting learning, teaching, and research. So we have adopted institutional repositories as a way to disseminate, store and preserve the information in order to allow permanent and reliable access of scientific production. It has been widely adopted in universities in national and international levels to assist in promoting access to information in a transparent and democratic manner, increasing the impact of recognition and national and international visibility of scientific production developed in the institutions \cite{medeiros2012}.

In this context, systems are identified that meet similar demands and have storage structures with similar data. In institutional repositories, the library management systems store in its database the metadata that contains information such as book title, author name, keywords that classify the content of a book, among others. These systems store a record of all activities performed by its users and can be useful to try to identify the profile and interest of each user who entered in the academic environment. However, it is essential that there is an integration capable of facilitating the search for content and patterned simultaneously in both systems.

The expansion of books databases and digital documents in libraries contributed to the increase in diversity of subjects, resulting in problems such as information overload. Usually, the user does not have perception of all content that is available for access, or only search content that is of their knowledge, but do not exploit other available subjects that may be of their interest.

In order to assist users in searching for relevant information, researches addressed similar issues, as the case of social tagging systems, where users express their preference sharing the tagging. Tags can be considered short words that convey some information and are used to identify the users interests, serving as a mechanism to generate personalized recommendations \cite{durao2012}. A recommendation system can combine this information to assist users in the effective identification of items suitable for a user’s need or preference, guiding them in a personalized way to suggest relevant items among a lot of options. Thus, it avoids the overhead of irrelevant information when it filters the access to information of users who have no detailed knowledge of what they want to find. Generally, these systems collect historical information produced over time by certain users, to try to trace your profile based on your preferences \cite{Barragans-Martinez2010}.

The techniques applied in recommendation systems are basically classified into two forms: collaborative filtering and filtering based on content. Collaborative filtering (CF) considers a group of users to make recommendations according to existing interests and certain similarity within the group. For this technique to be applied, a user must evaluate certain item. If there is any similarity among users, an item that has been evaluated can be recommended and it does not need to be evaluated by the user who will receive the recommendation \cite{gama2010}. In contrast, content based filtering technique depends on the user to evaluate an item to receive similar recommendations to the rated item, that way there is no dependence on the preferences of similar users and the recommendations are directed specifically to a user according to similarity of items of their interest \cite{gama2010}\cite{liu2015}.

The two recommended techniques described can assist users in choosing certain content, but they still have weaknesses that can be unfavorable to the recommendation process. The main disadvantage of the content based filtering system is related to the excess of expertise of a user in relation to an item, it means that a user tends to always choose items similar to what was chosen before. Items that do not have the features within the choice of a user patterns can be discarded, but if they were recommended as unexpected items could arouse the interest of the user \cite{kim2010}.

Collaborative Filtering approach also has some limitations, such problems related to sparse data, cold start and scalability \cite{polatidis2016} \cite{ji2015} \cite{durao2012} \cite{kim2010}. The sparse data problem occurs when the available data are insufficient to identify users or similar items, due to large number of users and items. This can occur when many users classify few items, or very popular items were classified by few users. Although it is possible to calculate the similarity, this does not become reliable because of insufficient information to be processed \cite{polatidis2016}. 

The cold start is a problem caused by the entry of new users in the system that have not yet conducted any type of evaluation which prevents the generation of recommendations until an item is classified.\cite{polatidis2016} divides this problem in cold start items when it comes to a new item that has not yet been evaluated, and the cold start user, when inserted new users who have not evaluated itens. The lack of scalability is a limitation that can affect the quality of the recommendations, in this case the number of users and items increases over time, making the recommendations impossible for new items and users, until there are ratings and evaluations among them \cite{durao2012}.

These problems presented can be alleviated by combining the FC techniques with the content based, resulting in an approach known as hybrid. Using this approach, the two techniques are combined to improve the accuracy of the recommendations, reducing the disadvantages and increasing the benefits \cite{liu2015}. When there is no information of users and their evaluations, mechanisms used in filtering based content systems are applied. If there is not enough information about the content associated with items, then the techniques used in collaborative filtering systems is applied \cite{kardan2013}. 

The use of these techniques in environments where users change their preferences constantly is not sufficient to ensure a suitable recommendation system. In a university environment, the academic community has more dynamic profiles, mainly the students, which have to search for a diversity of themes periodically. Therefore, we need an approach to adapt the recommendations according to new circumstances. To try to alleviate this problem, the concept of novelty detection are applied \cite{gama2010}, which makes it possible to recognize a concept as new and indicate the emergence of new concepts according to changes over time.

In this work, we intend to minimize the effort to search and improve the user experience by applying content recommendation on books databases and digital documents (IR) in a targeted and integrated manner. We implemented a hybrid recommendation strategy based on tags. For this, we extract information from books metadata that are part of the history of loans, and transform this information into lists tags that express the user's interest and represent their preferences. We apply the concept of novelty detection to adapt the recommendations to the user profile according to the changes of interest at a specific time. 

Users receive recommendations as a list of books and documents to evaluate, and they can return  their feedback. With this feedback, the system restructures the lists of tags according to the relevance of the evaluations, and generates new recommendations after these adjustments. To demonstrate the viability of this proposal, the developed approach is evaluated by accurately measures, recovery and f-score. So, we build a mechanism to provide recommendations directed to the user profile, by increasing the quality of their research in repositories and on the basis of library systems. In addition, we encourage the use of the open digital institutional repositories by providing greater visibility of the research developed by the institutions that are stored in these databases.

The rest of this paper is organized as follows. In Section 2, the related works are presented, the techniques and approaches that have been used, with their strengths and limitations. Section 3 describes the method used to collect the data used in the research. In Section 4 we describe how the system architecture was developed. Section 5 describes the methodology that was adopted to conduct the research. In section 6, the system implementation and evaluation of the experiments are described in detail. Finally, Section 7 concludes and points out directions for future work.

\section{Related Work}

Recommendation systems help users to effectively identify items according to your interest and need. Within a wide range of options, this system guides users in a personalized way to access content that is more relevant and useful. This type of system is used to reduce the information overload \cite{tejeda-Lorente2014}, being applied in several ways, as in e-commerce sites to improve sales, systems to indicate movies and TV programs, in blogs for articles recommendation, networks and social media \cite{Barragans-Martinez2010} \cite{gama2010} \cite{durao2012} \cite{liu2015}.

The work of \cite{tejeda-Lorente2014} proposes a recommendation system based on the quality of the resources. The system uses the quality of articles to estimate their relevance and apply an approach based on fuzzy logic, where users provide their preferences through the fuzzy linguistic concept to build your profile. This strategy applies a simple approach based on content, but it predicts the use of hybrid recommendation application to adapt collaborative filtering from the experience of recommendations shared among users. Furthermore, a module of the reclassification of content is incorporated, which combines the estimated relevance of an item with its quality. It is considered that the resources more preferable by the users are the ones that have good quality. Thus, it tends to generate more useful and accurate recommendations.

The study done by \cite{lai2013} adds the reliability of users in collaborative filtering technique to try to improve the quality of the recommendations and develop a hybrid trust model. The system calculates the value of the accuracy according to user evaluations on items already classifieds. In this case, a user is considered high confidence when contributes with more accurate classifications than others. Users who have some similarity are identified and the system forms groups to share items among its members. So, the user preferences affect the group, increasing personal confidence, which can change the recommendations from the group's perspective.

However, the confidence values may not have adequate accuracy when the number of items classified by a user is low. Therefore, it is necessary to use a hybrid trust model that considers all personal and group classifications. We used the tf-idf approach to determine the importance of the documents. A document that you download or upload is considered more important than the one that was visited. Thus, according to the access, the profile of users is being made. From that moment, the similarity among users is calculated by the cosine and the groups are formed based on their proximity values. The documents having a high forecast index are listed in a recommendation list. This method also improves the forecasting process because it uses the similarity among users to make recommendations.

The research by \cite{moreno2016} proposes a complete structure to deal with the most current problems in recommendation systems: scalability, dispersion and cold start. This work addresses the context of movie recommendation, but can be applied in other areas. The proposal is to combine web mining methods and ontologies to try to induce models on two levels of abstraction. The lower level model is built from data that do not use semantic information, while the higher-level model uses web classified data with semantic information according to defined ontology. With the combination of these models, patterns are generated with a high level of abstraction, able to relate types of products and user profiles in a widespread way. Thus, the model is able to recommend products that have not been evaluated or make recommendations for new users. Moreover, the model is able to deal with the dispersion problem using associative classification methods.

In the literature we also find researches that address the use of tags to improve the quality of recommendations, such as the study done by \cite{kim2010} which uses collaborative filtering combined with collaborative tagging to enhance recommendations using tags created by users. In this research, the approach known as collaborative tagging creates user profiles according to the markings made on the items. This is done in two phases: one phase of models elaboration and the other of probabilistic recommendation. Using a collaborative filtering scheme, the markings are generated (set of candidate tags) that may be directed to a user. Through these tags, the Naive Bayes algorithm is applied to recommend items. The marking system consists of the relation of three elements: user, item and term. These three elements are analyzed considering their frequency of a binary vector, via the user-item relationship, user-tag and tag-item where the user makes a mark on an item using a term. Through the collaborative filtering process the system finds other markings similar to the tags of a candidate tags group and the recommendations for users according to the similarity of personal tags for each user.

The data used in this study were collected from the site del.icio.us, which is a social tagging service that uses collaborative tagging. The research identified that the proposed algorithm achieved significant results regarding the quality of the recommendations, in addition to reduced problems related to sparse data and cold start. It was also observed that the most suitable method provided the user profile items, even if the amount of recommended items was smaller. It was identified problems related to polysemy and synonymy, which affects the quality of the set of tags, but it is a limitation that can be fixed in the future by the application of semantics.

Other scientific research using keywords (tags) have been adopted in other studies applied to recommendation content. A number of articles have been developed \cite{durao2009} \cite{durao2010} \cite{dolog2011} \cite{durao2012} \cite{durao2014}, proposing to improve the qua-
lity of recommendations at every stage of research by combining methods applied to hybrid content recommendation, tags and collaborative filtering. In these studies, hypertext pages as Wikis are raided to search tags (keywords) that may represent some content, that identify and transmit some meaning on the page. Based on user access to links on a page that redirects to another, the choices will be accounted for in accordance with the word that served as anchor to access the new page. Markings made by users are a way to classify and evaluate the page, and makes possible the creation of the profile that allows the direction of suggestions to be included on the page.

It is noticeable that most studies consider the limitations presented by the recommendation techniques, and therefore adopt a hybrid approach and increase other resources such as the formation of tags to match more than one technique and try to address these limitations.

\section{Gathering Information and Feedback}

An important step for the formation of a user profile on a recommendation system is the process of gathering information, which can usually be done in two ways. In explicit form which the information are absorbed by the system when a particular user enters their preferences, and the implicit form, that uses behavioral analysis, by capturing the paths (links) driven by a user on a website.

Explicit information is collected when users make evaluations and indicate their preferences and interests, which can be collected through pre-filled forms. The discovery of implicit information is a more costly process, because depending on the field of application information may be hidden. Some methods are used to extract implicit information from the availability of user data, such as: activities and behavior, existing relationships among users, mapping of the items that were visited, and time spent watching an item \cite{gama2010}.

After the first recommendations, it is important to know the opinion (feedback) from users on the relevance of what has been suggested to improve the accuracy of the system in future indications content. Feedback can be provided implicitly and explicitly. In the explicit form the users return their opinions through ratings, textual comments or inform in binary scale or show no interest in the content. When implicit feedback is applied, automatic inference techniques monitor the actions of users to discover their preferences. To make this possible, it uses the browsing history of the user, links that are accessed and time spent on a page. The combination of the two methods to assess whether the identified user behavior implicitly is according to their assessments retrieved explicitly. Thus, a hybrid alternative is constructed that is able to verify if the user acts follow a logical, in other words, if the acts are consistent in their patterns of interests \cite{reatengui2006}.

\section{System Architecture}

Some studies in the literature use a hybrid approach, combining collaborative filtering and filtering based on content. In our work the hybrid approach is applied and enhanced to use tags. Furthermore, it adopts the concept of novelty detection to identify changes over time of interest to users.

\begin{figure} [!htb]
	\centering
	\includegraphics[width=\textwidth,keepaspectratio]{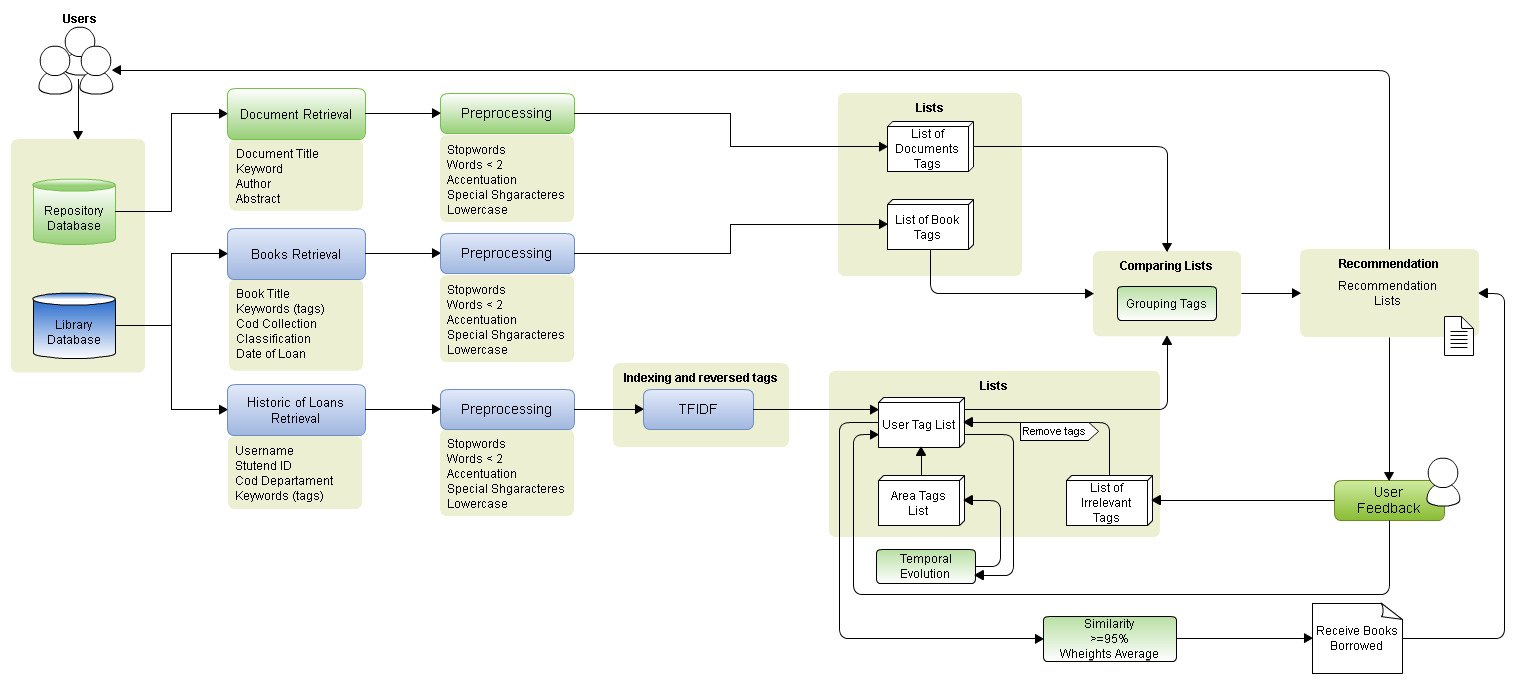}
	\caption{\textit{System Architecture}}
	\label{fig1}
\end{figure}

The user profile is formed by the tags that are retrieved from the metadata of the books borrowed by users, so each user will have a list of tags. The concept of content based filtering is applied to get the book loans history of library users, which is combined with a list of books recommendations sent to the user to evaluate each item with a note. Collaborative filtering is used to identify the proximity among the user profiles and to recommend content among those who have high similarity of interests. The novelty detection concept is applied to identify the changes in the list of tags of each period of a particular course. In figure \ref{fig1}, is presented in detail the system structure and its stages, which will be detailed in the next topics.

\subsection{Collecting Information Loans}
In figure \ref{fig1} there are two databases, one referring to the library management system and another representing the institutional repository. Institutional repository is stored digital documents such as theses, dissertations, and articles. The library database stores all information necessary for management of academic activities, and settled daily through actions of users (students, faculty, and staff). When a user loan a book, a lot of information about this action are stored in the database. This information contains metadata that will be used to extract tags and map the user profile. Metadata is formed by the following attributes:

Metadata of the users loans history:
\begin{itemize}
    \item User name: user name who made the book loan
    \item Student ID: unique numerical value that identifies a student in the system
    \item Cod Department: the department code to which the student who made the loan is linked
    \item Keyword (tags): This metadata is related to information in the book that was borrowed by the user. This information is manipulated and processed into tags. 
\end{itemize}

Books metadata:
\begin{itemize}
\item Book Title: corresponds to the name of the book
\item Keywords (tags): This field stores information as book author, area and keywords that characterize the content of the book.
\item Cod Collection: unique numerical value that identifies a particular book 
\item Classification: code that classifies a book according to the area of concentration
\item Date of loan: date referring to the book of the loan made by the user
\end{itemize}

This information is used to generate the list of tags that form each user's profile. To better get user preferences with more accurate information, a list of books is sent as a recommendation, which is evaluated with a note for each item that was recommended. Thus, the rating impact on the user list tags and improve the quality of recommendations. 

The information from the repository database documents is integrated with the library database to generate recommendations from both databases. In addition, to being an alternative to complement the variety of information, it also contributes to promoting the use of the repository, since users will have a single interface to access content from both systems. The attributes used to extract information from the digital document metadata are:

\begin{itemize}
\item Document title: title of the document stored in the repository
\item Keywords: terms that are used as key words of the document
\item Abstract: textual description used to describe the subject covered in the document
\item Author: identifies the people who developed the work
\end{itemize}

\subsection{Preprocessing of the Data Collected}

At this stage, the data were collected from the database will be filtered to keep only what is relevant to the search. So, we apply techniques of selection and processing of data used in most systems that work with data mining or retrieval information. 

First, the stopwords are removed because even if used to form a sentence, rarely contribute to add meaning to the words. The stopwords are considered words that occur with a high frequency in the documents, but potentially become useless because they are very common, not contributing to the relevance of the research. Examples of such words are: the, and, to where, that, was, this, among others. A stopwords file is formed to be used by the system to remove from the information retrieved, all existing words in this list \cite{elmasri2011}.

Other removals are made to maintain the quality of the information retrieved. Words with less than two characters are disregarded, and to facilitate the analysis are removed accents, special characters, and all the words are converted to lowercase.

In the first test is applied stemming process, also known as stemming technique. This process is used in many computer applications to transform the variant forms of a word in a more accurate representation, and that is generic enough to capture the essence of the words \cite{alvares2014}. The approach used in stemming process applied in this study was based on Snowball algorithm, a variation of Porter's algorithm, which uses the removal of suffixes \cite{porter2016} \cite{xavier2013}. However, application of this technique was not satisfactory for this experiment. When using stemming algorithm accuracy decreases. A possible cause may be the creation of keywords (tags). In this context the words are not created freely by the user, but by specialists who analyze the terms before registering the information in document metadata. Therefore, considering the loss of accuracy and also the demand for computational processing to perform this task, the stemming utilization in this application was not continued.

\subsection{Tags List}

Some studies adopt a model to represent features in documents, known in Information Retrieval as Bag of words \cite{jiang2010} \cite{luo2011} \cite{lin2016}, and has been considered promising in the content classification. In this model, any textual content is transformed into a set of words that convey some meaning and are used as the basis for classification of a given resource. Following this concept, we adapt the model for our study to use the tags list (bag of tags).

The tags lists are formed by a set of words extracted from books metadata that are collected in two ways: record of the books that were borrowed by users in the databases and lists (feedback) of recommended books for the user. Each term or word is considered a tag that expresses a meaning that classifies the content to which this tag was extracted. In this case, if a user is borrowing a book whose title is ``Computer Programming", the extracted tags of this book metadata represent its context, and therefore it indicates that the user is interested in documents relating to this subject. Each user profile is traced to recommend content, based on their tags set.

\begin{figure} [!htb]
	\centering
	\includegraphics[width=\textwidth,keepaspectratio]{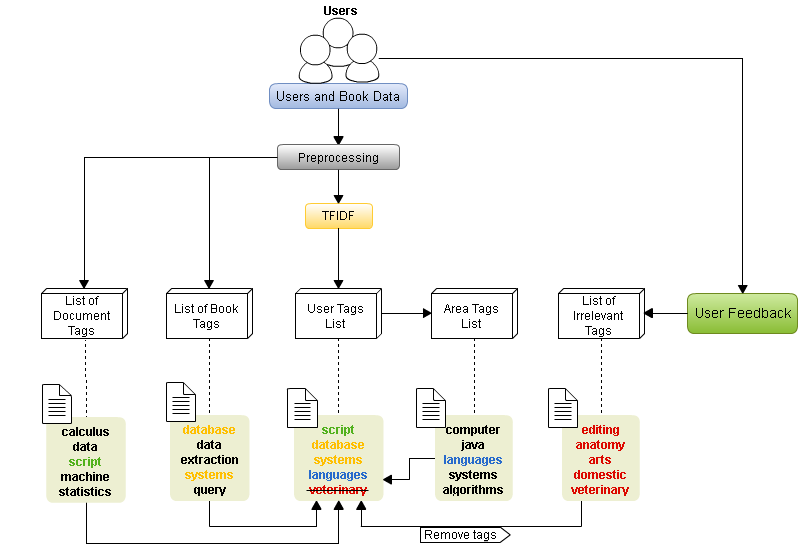}
	\caption{\textit{Tags list and their formation.}}
	\label{fig2}
\end{figure}

In this work, we implemented a method that consists of five lists of tags that are applied during the recommendation process. This phase is illustrated in figure \ref{fig1}, but to facilitate the understanding it is represented in more details in figure \ref{fig2}.

The five lists of tags shown in figure \ref{fig2} are: list of document tags, list of books tags, list of user tags, list of area tags, and lists of irrelevant tags. With the exception of the list of document tags and books, the other three can be considered lists of dynamic tags because they are changed over time according to the changes of user's interest.

\subsubsection{List of Books Tags}

\begin{figure} [!htb]
	\centering
	\includegraphics[width=\textwidth,keepaspectratio]{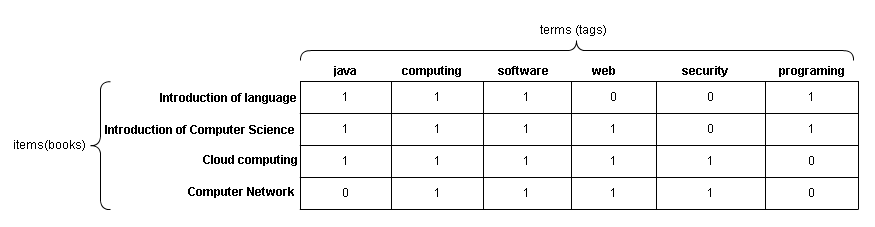}
	\caption{\textit{Books vector x tags.}}
	\label{fig3}
\end{figure}

This list is formed from the extraction of metadata to classify a particular book. Through a connection to the library database the information of the books contained in the system database is captured. A query returns the book metadata that are extracted and processed in the preprocessing phase, forming the lists of books tags. This list is immutable, in other words, their tags are always the same for each book unless its metadata is changed in the system. This list is represented in a binary vector of items $x$ terms ($i$, $t$), where i refers to the item, which in this case is the recovered book and t refers to the term which is the word that is extracted from the metadata, which we call tag. The figure \ref{fig3} illustrates how the structure of this list.

The vector that represents the relationship items-terms (books-tags) is formed by the presence or not of a certain tag in a book. If the book has the tag then receives value 1, otherwise the value is 0.

\subsubsection{List of Documents Tags}
The formation of this list follows the same principle of book list tags. However, the metadata are extracted from documents stored in the institutional repository. A query is used to return the document metadata that are extracted and processed in the preprocessing phase, forming the lists of document tags. This list is also immutable, in other words, the tags are always the same for each document, unless its metadata is changed.

\subsubsection{User Tag List}

The user tag list is formed by the words extracted from the metadata of the books in the library database, considering the user loan history in a certain period of time. This list interacts with the list of irrelevant tags and area tags, which suffer changes over time and it adapts the tags according to user behaviors in the system. A list of such flux can be observed in figure \ref{fig2}. 

After this step, the tags are generated and each of them is set a weight. The tags that have a higher weight value are considered the most important in the user list, as they represent high user interest for the content expressed by the tag. The statistical method TF-IDF was adapted in this study to assign weight to the tags. This method is used to discover the importance of words in unstructured text or semi structured. In the methodology section the TF-IDF method and how the adaptation was made to be applied in this study will be explained in more detail. 

The user tag list is represented in a binary vector users $x$ terms ($u$, $t$), where u refers to the user, which in this case is the person who performs loan transactions in the database and $t$ refers to the term, which is the word extracted from the metadata of the books that were borrowed. An example is shown below in figure \ref{fig4}.


\begin{figure} [!htb]
	\centering
	\includegraphics[width=\textwidth,keepaspectratio]{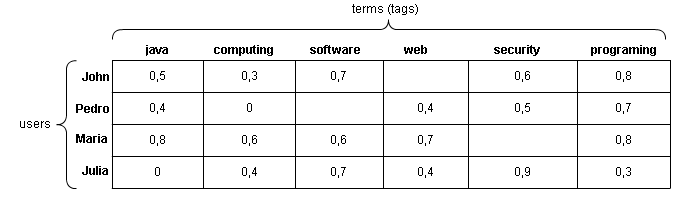}
	\caption{\textit{Users vectors x tags.}}
	\label{fig4}
\end{figure}

In figure \ref{fig4} each tag of a particular user has a weight. A high weight value represents a greater importance term degree for the user. The user John has the tags software and programming with more weight, which means that this user has more interest in subjects related to contents expressed by the tags meanings. The terms of the vector that have no values, are not part of the user tags list. This can occur if the users lent books do not have these terms.

\subsubsection{List of Irrelevant Tags}

Systems designed to recommend content, as well as other types of systems, are always subject to errors, because not always an item that is recommended for the user is according to their interest. When a user is not interested in a recommended item, this item is classified as a false positive, in other words, it was suggested, but not classified as an item of their interest, and therefore should not have been recommended.
 
This type of problem can be compounded in an environment such as the libraries. Some users do book loans to transfer them to others. This can decrease the accuracy of the recommendation system because the user has no interest in the loan book, which generates an incompatible suggested content with the user's needs.

To alleviate this problem, we implemented the list of irrelevant tags to identify terms that are not relevant. Initially, the user receives a list of recommended books based on their tags and evaluates each recommendation with a score from 0 to 3. A close evaluation of 3 indicates greater user interest for the recommended item. When a book is classified by the user with a score of 0, then this item is not the user’s interest. Thus, an irrelevant recommendation is identified (false positive) by the user feedback to form the list of irrelevant tags. Each user will have a list of irrelevant tags, which will be formed by the tags of the books that are classified as grade 0 in the recommendation list. 

\begin{figure} [!htb]
	\centering
	\includegraphics[width=\textwidth,keepaspectratio]{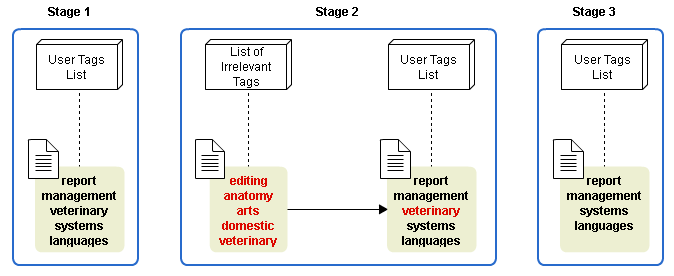}
	\caption{\textit{Irrelevant tag removal flow.}}
	\label{fig5}
\end{figure}

Then, when new recommendations are made, the user tag list changes based on irrelevance list. Tags classified as irrelevant are removed from the user's tags list to avoid recommendations on terms that do not represent user interest content. As an example, observing figure \ref{fig2}, in step 1 to the user tags list contains terms extracted from the borrowed books metadata. When the user evaluates an item with note 0 through feedback, is formed in step 2, the ``List of Irrelevant Tags" that contains the tags: editing, anatomy, arts, domestic, and veterinary. The tag ``veterinary" as irrelevant is detected and removed from the ``User Tag List". From that moment, the ``Tags User List" (step 3) no longer contains the tags responsible for generating recommendations that are not relevant to the user. The tag removal flux is shown in figure \ref{fig5}. 


\subsubsection{Area Tags List}
The content recommendation systems have limitations that may impair its effectiveness. When new users access the system, evaluation items are usually not made initially, which makes the content recommendation process and the measurement of the similarity degree among users. This can be a recurring problem in libraries, because some users do not usually offer book loans for some time, making it difficult to identify their interests. The current course transfers also impact on changes of interest, since the subjects and themes are different according to each period of the course. Even if the user does not change the course over the period, the subjects have a different content, which may change the interest and importance on the contents.
 
To alleviate this problem, the list of area tags is used, which is formed by the tags extraction of the borrowed books in the course period. Thus, we identified the course profile at a certain moment in time. Each course will have a list of dynamic tags for each period that may change according to the loans made by users who are enrolled in that period. When a user enters in a course the tags related to the course period loans are directed to the user tags list, allowing the period content to be suggested to the new users.

As an example, suppose Alice, John and Charles are enrolled in the third period of the Management course and they loan many books. Based on these loans, the course tags list in that period is created, and it will serve to indicate content for Marcos who will attend the third period of Administration. This makes Marcos to recieve content recommendations that are generally used in the period in which he is enrolled. This can assist in Marcos choices, as in previous years he had not made loans, having not on his list any tags that could represent his preferences. However, based on the area tags, it was possible to generate recommendations on subjects that will be used in the period and that can probably be of his interest.

As is done in the user's list tags, the tags of the area list are also given a weight that is calculated by the statistical method TF-IDF to identify the relevance of the tag in the list.

\subsection{Grouping Tags}
At this stage the tag lists are compared to generate recommendations. We consider three threshold groups defined based on the weight of the tags: 

\begin{itemize}
\item Group 1: $\geq 70\%$;
\item Group 2: $\geq 40\%$ and $\leq 69\%$ ;
\item Group 3: $\leq 39\%$.
\end{itemize}

The tag belongs to group 1 when its weight is 70 higher than the others. For a recommendation is based on that group, at least four tags extracted from a book or document must be identical to the group's tags 1. If this condition is not satisfied, the extracted tag will be analyzed based on the group 2, in this case at least 5 tags must be compatible. Also, in group 3 at least 5 tags that are compared must be identical to have some recommendation. The amount of tags and percentage of groups were defined observing the precision and recall values, obtaining better results in the experiments. So, the algorithm was run repeatedly varying the percentage criteria of groups and number of tags to identify the best cutoff values.

\subsection{Obtaining the User’s Profile}
Two strategies are used to trace the user profile: mapping of the historical loans and collecting feedback. First, the borrowed books metadata are extracted to form a list of tags that will represent their interests. Thus, each user has a list of tags that are used to generate recommendations of books and documents.

In the second stage, a list of items consisting of books and documents is submitted to the users, then, they send feedback of each item according to their interest. The user tags list is changed according to the evaluations. Thus, when the system combines information from the data collected by the feedback and the record loans, it becomes possible to obtain a complete and reliable profile according to the user interest.

\begin{figure} [!htb]
	\centering
	\includegraphics[width=\textwidth,keepaspectratio]{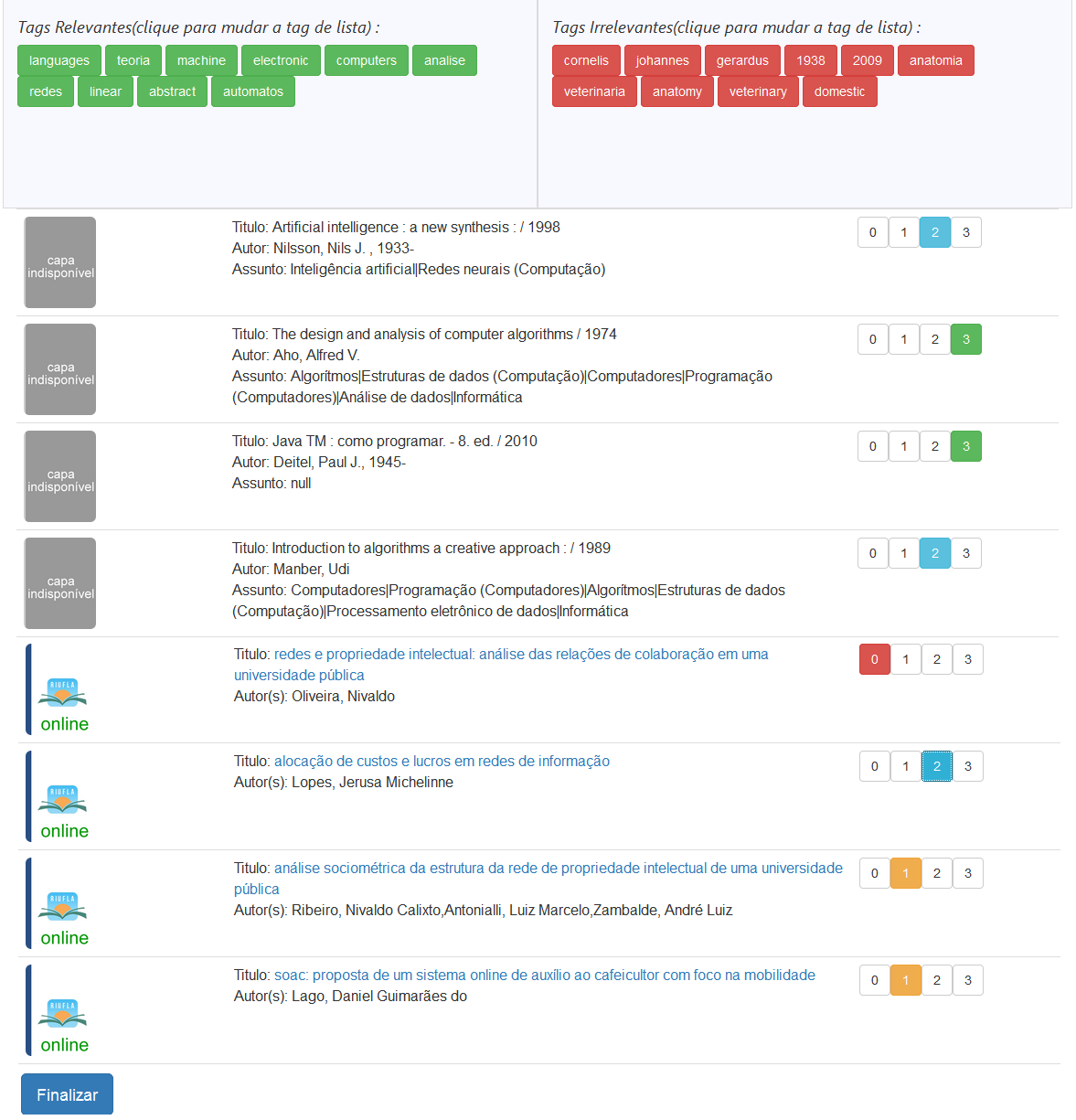}
	\caption{\textit{User’s feedback screen.}}
	\label{fig6}
\end{figure}

The figure \ref{fig6} shows the screen with the recommendations that the user must evaluate. Each user has a list of personalized recommendations. The access to this information is done through a link that is sent to the email of each user who received the recommendations. The message has a link to access their list of recommendations and information that should be considered to classify items. In this case, the user must evaluate each item a score from 0 to 3, where:
\begin{itemize}
\item Value 0, means that the recommended item does not interest to the user;
\item Value 1, recommended item has a little interest to the user;
\item Value 2, recommended item interests to the user;
\item Value 3, the user has much interest in the recommended item.
\end{itemize}

If the item is rated with a score 0, we consider that the recommendation should not have been suggested to the user. Thus, we extract the metadata for that item and turn it into tags that are sent to the list of irrelevant user tags. The next recommendation items that have some relationship with these irrelevant tags, will no longer be displayed to the user. At the top of the page, the user has access to a list of relevant and irrelevant tags, and can reallocate them among the lists according to their interest. If the document is recommended from the repository, by clicking on its title, the user is redirected to the page that contains complete information describing in detail the contents of the item to be classified.

\section{Methodology}
The following topics we explain the methods of content based filtering and collaborative filtering that have been used and adapted for this study.

\subsection{Content Based Filtering}
The content based filtering approach identifies the user preferably by similar items, which were previously classified as interest. Based on information and specific data about a user, it is possible to identify whether there is any relationship between the user and the content \cite{valois2013}. In our research we considered that the user's interest is related to the books that were borrowed. Thus, we can identify through the information contained in the books metadata which subjects are interesting to them. 

To perform the content filtering, we use the space vector model. With this model it is possible to select an array of items that most resembles a key item. When selecting a desired item, it calculates the similarity among each of the items of these vectors based on the key item \cite{Barragans-Martinez2010}. In our model, we apply some concepts of this technique to collect information that is important to designate the user's interest. We extract the textual data of the books’ metadata to generate tags and use the statistical measure TF-IDF (Term Frequency - Inverse Document Frequency) to establish the importance of weights for each tag. 

The TF-IDF technique is used in text mining to identify the importance of words in a text. A value is assigned to each term extracted from the text according to the word frequency in the text or in the documents. This value represents the weight that determines the importance of the term in the text, and all documents in the database. The equation (\ref{eq1}) shows how is the TF calculation \cite{elmasri2011}: 

\begin{equation}
    TF_{ij} = \frac{f_{ij}}{\sum_{i=1}^{|V|}f_{ij}}
    \label{eq1}
\end{equation}

In this equation the symbols have the following meaning:

\begin{itemize}
\item $TF_{i,j}$ is the frequency of the standard term to term $i$ in the document $D_i$.
\item $f_{i,j}$ is the number of occurrences of the word $i$ in document $D_i$
\end{itemize}

The IDF calculation is done as shown in equation (\ref{eq2}):

\begin{equation}
    IDF_{i} = \log(\frac{N}{n_{i}})
    \label{eq2}
\end{equation}

Where:

\begin{itemize}
\item $IDF$ is the inverse of the frequency weight of the document to the term $i$.
\item $N$ is the number of documents in the collection.
\item  $n_i$ is the number of documents in which the term $i$ occurs.
\end{itemize}

The TF-IDF uses the product of the normalized frequency of a term $i$ ($TF_{ij}$) in the document $D_i$ and the inverse frequency of the term of the document $i$ for determining the weight of a term in the document. The calculation is represented in equation (\ref{eq3}):

\begin{equation}
    TF - IDF_{ij} = TF_{ij}\times IDF_{ij} 
    \label{eq3}
\end{equation}

The frequency utilization only can be misleading because not always the words that are most common are the most important. Therefore, calculating the weight TF-IDF can be useful to determine a more reliable value of relevance of the term in the document. 

We adapt this calculation to apply on our data and we consider that a book borrowed by a user is a document and the terms are information extracted from their metadata, such as title, keyword, and author. So, many books represent a set of documents. To calculate the TF-IDF we consider that documents are equivalent to books and their metadata equivalent to terms that are part of a document.

\begin{equation}
    TF_{tl} = \frac{f_{tl}}{\sum_{i=1}^{|V|}f_{tl}} 
    \label{eq4}
\end{equation}

where:
\begin{itemize}
\item $TF_{tl}$ is the frequency of the term normalized to term $t$ in the book $L_i$ .
\item $f_{tl}$ is the number of occurrences of the term $t$ in the book $L_i$.
\end{itemize}

To calculate the IDF represented in equation (\ref{eq5}) we consider:

\begin{equation}
    IDF_{t} = \log(\frac{N}{n_{t}})
    \label{eq5}
\end{equation}

where:
\begin{itemize}
\item $IDF_t$ it is the inverse of the frequency weight of the book for the term $t$.
\item $N$ is the amount of whole books.
\item $n_t$ is the number of books in which the term $t$ occurs.
\end{itemize}

Thus we calculate the TF-IDF in equation (\ref{eq6}):

\begin{equation}
    TF - IDF_{tl} = TF_{tl}\times IDF_{tl}
    \label{eq6}
\end{equation}

Describing the adjustment of the formula, the TF is the fraction between the number of times a word occurs in all borrowed books by the total number of terms of the whole books. And the IDF is the fraction of the total number of books by the amount of books that a term occurs. Considering that the user Miguel lent the books ``Java Programming", ``Introduction to C++", and ``Computer Networks". We extract the metadata of these books and get the following information for the books, as table \ref{tab1}:

\begin{table}[]
\centering
\begin{tabular}{|l|l|l|l|}
\hline
\textbf{Books}      & \multicolumn{3}{l|}{\textbf{Terms}} \\ \hline
Java Programming    & programming  & oriented  & objects  \\ \hline
Introduction to C++ & programming  & objects   & C++      \\ \hline
Computer Network    & tcpip        & layers    & security \\ \hline
\end{tabular}
\caption{\textit{Books and terms (tags).}}
\label{tab1}
\end{table}

We use the calculation of the TF-IDF to obtain the weight of each term. The term ``object" would have the following results:

\begin{center}
 $  TF_{tl} = \frac{2}{9} = 0.222$

  $  IDF_{t} = \log(\frac{3}{2}) = 1,5 $
    

 $   TF - IDF_{tl} = 0,222 \times 1,5 = 0,333$

\end{center}

In TF calculation, the value 2 is the number of occurrences of the term ``object" in any set of terms. The value 9 represents the total number of terms in the set. In the calculation of the IDF, the value 3 is the amount of books of the set, and the value 2 is the number of books in which the term ``object" occurs. As a result of TF-IDF calculation we obtain the weight value of 0,333 for the term ``objects".

We adapt the calculation because it usually determines the importance of the term in the document, but in this case, we identify the importance of the term for a particular user. Then, we fill in the user tags list with the relevant weights found for each term.

\subsection{Collaborative Filtering Approach}

This technique generates suggestions for particular user opinions and considering similar interests of other users, according to their characteristics or similar behaviors \cite{kardan2013}. There are several ways to calculate the similarity, the most used are cosine, Pearson correlation and mean squared difference (MSD). We adopted the similarity cosine because it has been used satisfactorily in several studies \cite{usharani2013} \cite{oliveira2013} \cite{durao2010} \cite{kim2010} and \cite{chirita2007}.

A comparative study on similarity metrics was done by \cite{salazar2012} in which two sets of data were integrated, one containing 574 scientific articles and other containing 1771 documents, collected on the web. The metrics were compared and verified for each of the processing time in relation to the size of the string. The metric that had the best time was the similarity of cosine and overlap coefficient, while the SmithWaterman required more time. In equation (\ref{eq10}) shows the cosine similarity calculation.

\begin{equation}
   similarity = \cos(x,y)=\frac{A\times B}{|A||B|} = \frac{\sum_{i=1}^{n}A_{i}B_{i}}{\sqrt{\sum_{i=1}^{n}A_i^2}{\sqrt{\sum_{i=1}^{n}B_i^2}}}
    \label{eq10}
\end{equation}

where $A_i$  e $B_i$ are vector components $A$ and $B$, respectively. The value generated by calculating the similarity cosine is in the range [0,1], where in 1 indicates an exact match between vectors, while 0 indicates the opposite. Therefore, the closer to 1 the result is, greater is the degree of similarity \cite{zhu2011}.

An important point that should be taken into account in applying this metric is the choice of the amount of similar users to generate recommendations. A very small amount of users may not be enough to generate consistent rules, because the likelihood of these users to classify an item that an individual has an interest is small. However, a high number of users can generalize the rules, due to higher proportion of users who do not resemble the characteristics of others \cite{sanchez2008}.

In this study we used the calculation of the cosine similarity combined with the TF-IDF calculation to identify users with similar preferences according to the weight of each user tag. Thus, we recommend books that are borrowed by a user, but that may interest the other according to a degree of similarity between them. The calculation of similarity that was used  follows the equation (\ref{eq11}):

\begin{equation}
    similarity(U_{x},U_{y}) = \frac{\sum_{i=1}^{n}U_{xt}U_{yt}}{\sqrt{\sum_{i=1}^{n}U_{xt}^2}{\sqrt{\sum_{i=1}^{n}U_{yt}^2}}}
    \label{eq11}
\end{equation}

where $U_{x,t}$  is the user tag weight $x$ and $U_{y,t}$ is the user tag weight $y$. By obtaining the similarities between users, an array $U_x \times U_y$  is formed with values that identify the degree of similarity among users.

To identify the proximity between two users based on their list of tags, consider the following hypothetical situation. User A has in its list the tag X with weight 0.3, the tag Y with weight 0.0 and the tag Z with weight 0.5, while the user B has the tag X with weight 0.5, the tag Y with weight 0.4 and the tag Z with weight 0.3. Each of the tags is part of the user tags list with the weight calculated by the TF-IDF. Applying the values in the similarity formula we have the following results:

\begin{equation}
    similarity(U_{1},U_{2}) = \frac{(0.3\times 0.5)+(0.0\times 0.4)+(0.5\times 0.3)}{\sqrt{0.3^{2}+0.0^{2}+0.5^{2}}\times{\sqrt{0.5^{2}+0.4^{2}+0.3^{2}}}}=0.73
    \label{eq12}
\end{equation}
    
The result of the similarity calculation between the user A and user B, resulted in a degree of 0.73 similarity between them. With the entire calculated results the matrix $U_1 \times U_1$ is created. 

\begin{table}[]
\centering
\begin{tabular}{|l|l|l|}
\hline
                & \textbf{User 1} & \textbf{User 2} \\ \hline
\textbf{User 1} & 1.0             & 0.73            \\ \hline
\textbf{User 2} & 0.73            & 1.0             \\ \hline
\end{tabular}
\caption{\textit{Matrix users.}}
\label{tab2}
\end{table}

We use these values similarity among users to indicate books which have a high degree of similarity. In our application, users who have a degree of similarity higher than or equal to 95$\%$ receive the books that were borrowed by them as a recommendation. It is considered a higher percentage to avoid recommendations among users who have little resemblance relationship, less likely to suggest books that are not appropriate to the user profile.

\section{Experiments and Evaluations}
In this section, we will present the proposal and evaluation of the recommendation system based on tags. It is verified if the user's system recommendations that were sent to the users are correct. To do this analysis, we compare the recommendations generated by the system with the history of user loans. In the next step, we verify if the system correctly predicts the ratings of recommended items, based on the users feedback made through the list of items that were suggested. For the methods evaluations precision measurements, recall, and f-score were used.

\subsection{Data Set}
To accomplish this research were collected information from the library management system database and the university's institutional repository database. The library database contains records of activities held at the library, as well as the history of loans, information about users and books. The data from the institutional repository are digital documents, containing information of articles, dissertations and theses.

To the library database were considered 27,767 loans from 01/01/2014 to 06/01/2015 period in which 1,769 users receive recommendations and 122 people sent their feedback. Many users use personal email, so do not access the link of recommendations list that is sent to the institutional email. In the self-evaluation institutional report developed by Self Evaluation Institutional Commission UFLA - CPA was found that only 54.7$\%$ of students use the institutional email \cite{scolfro2016}. The users lack of interest to participate and classify the recommendations also affects data collection.

\begin{table}[]
\centering
\begin{tabular}{|l|l|l|l|l|l|}
\hline
\multicolumn{3}{|l|}{\textbf{Library Database}} & \multicolumn{3}{l|}{\textbf{Repository Database}} \\ \hline
Books       & Tags         & Tags (filter)      & Documents      & Tags         & Tags (filter)     \\ \hline
1,795       & 988,848      & 363,392            & 9,478          & 693,089      & 495,259           \\ \hline
\end{tabular}
\caption{\textit{Library database information.}}
\label{tab3}
\end{table}

According to table \ref{tab3}, after the processing phase the amount of recovered tags from the library database decreases to 363,392 and the recovered tags repository after treatment reduce to 495,259, without considering repeated terms.

Users involved in the research are part of 6 different courses: Master in Computer Science, Undergraduation in Computer Science, Information Systems, Administration, Animal Science, and Veterinary Medicine. Purposely, to verify the consistency and to evaluate the algorithm recommendations with more criteria were selected courses that have different content. 

The list of recommendations contains 30 items including books and documents for each user to evaluate. A larger number of items can make the evaluation process expensive, causing dropouts at the time of classification.

\subsection{Recommendations Evaluations}

In recommendation systems, precision metrics, recall and F-Score, they are used to assess if the system recommends items that are actually considered relevant by the user. To enable the calculation of these measures is necessary to quantify and to categorize the items as their information \cite{tejeda-Lorente2014}. Items can be classified as relevant or irrelevant and recommended or not recommended, as table \ref{tab4}:

\begin{table}[]
\centering
\begin{tabular}{|l|l|l|}
\hline
                    & \textbf{Recommended} & \textbf{Not Recommended} \\ \hline
\textbf{Relevant}   & Nrr (TP)             & Nrn (FN)                 \\ \hline
\textbf{Irrelevant} & Nir (FP)             & Nin (TN)                 \\ \hline
\end{tabular}
\caption{\textit{Table relationships.}}
\label{tab4}
\end{table}

This table uses the same conception of the confusion matrix, where terms like true positive (TP), false positive (FP), false negative (FN), and true negative (VN) are applied to compare specific classification of an item with the classification correct desired, and their values are used in the calculation precision, recall and F-score.

Accuracy is defined by the ratio between the relevant items recommended by recommended items. This metric is used in order to measure the probability of a recommended product to be relevant to the user. The metric is defined in equation (\ref{eq13}):

\begin{equation}
    P = \frac{N_{rr}}{N_{rr}+N_{ir}}
    \label{eq13}
\end{equation}

Recall is calculated by the ratio of relevant items recommended by relevant items. Is the probability of a relevant item to be recommended. The metric is defined in equation (\ref{eq14}):

\begin{equation}
    R = \frac{N_{rr}}{N_{rr}+N_{rn}}
    \label{eq14}
\end{equation}

F-Score (F) is the harmonic mean combination of precision and recall values. The measure is used to compare different sets of results. The metric is defined in equation (\ref{eq15}):

\begin{equation}
    F = \frac{2\times R\times P}{R+P}
    \label{eq15}
\end{equation}

\subsection{Implementation and Evaluation of the Experiment}

In the following topics, algorithm execution example is described, through all the stages already described above. Then we evaluate the results based on the metrics mentioned in the previous section.

\subsection{Experiment Implementation}

To start the experiments, information from the library database and from the repository are integrated and collected for the user profile construction. The history of user loans in the library is used to generate the users profile from the books metadata extraction. The repository documents, as well as the books are recommended for evaluation in the feedback stage. 

In the preprocessing step, the information collected is subject to various filtering treatments to form the lists of tags that are used to generate recommendations. Applies to TF-IDF technique that sets the weights tags to identify the degree of importance of the term for the user. The tag lists of books and the list of documents tags are formed without weights because it is not necessary to identify the importance of the term in the document, but what its importance to the user. In table \ref{tab5} we present the set of borrowed books of a real user we call UserStudent:

\begin{table}[]
\centering
\begin{tabular}{|l|l|}
\hline
\textbf{User}                 & \textbf{Borrow books}                                                                                                    \\ \hline
\multirow{10}{*}{UserStudent} & INTRODUCAO A PROGRAMACAO LINEAR\\ \cline{2-2} 
                              & ORGANIZACAO ESTRUTURADA DE COMPUTADORES \\ \cline{2-2} 
                              & ELEMENTS OF THE THEORY OF COMPUTATION\\ \cline{2-2} 
                              & LINGUAGENS FORMAIS E AUTOMATOS \\ \cline{2-2} 
                              & PESQUISA OPERACIONAL                                                                                                     \\ \cline{2-2} 
                              & \begin{tabular}[c]{@{}l@{}}FUNDAMENTOS MATEMATICOS PARA A CIENCIA DA\\ COMPUTACAO\end{tabular}                           \\ \cline{2-2} 
                              & LINEAR PROGRAMMING AND NETWORK FLOWS                                                                                     \\ \cline{2-2} 
                              & INTRODUCTION TO ALGORITHMS                                                                                               \\ \cline{2-2} 
                              & \begin{tabular}[c]{@{}l@{}}TEXTBOOK OF VETERINARY ANATOMY FOURTH \\ EDITION TRATADO DE ANATOMIA VETERINARIA\end{tabular} \\ \cline{2-2} 
                              & ALGORITMOS E SEUS FUNDAMENTOS                                                                                            \\ \hline
\end{tabular}
\caption{\textit{Books borrowed by a user.}}
\label{tab5}
\end{table}

In table \ref{tab6} we present the set of tags which were taken from the metadata of these books being identified their importance weights defined from the application of TF- IDF calculation:

\begin{table}[]
\centering
\begin{tabular}{|l|l|l|}
\hline
\textbf{User}                 & \textbf{Tags of user} & \textbf{Weight} \\ \hline
\multirow{16}{*}{UserStudent} & logic                 & 0.0662          \\ \cline{2-3} 
                              & languages             & 0.0191          \\ \cline{2-3} 
                              & electronic            & 0.0514          \\ \cline{2-3} 
                              & computers             & 0.0410          \\ \cline{2-3} 
                              & teoria                & 0.0164          \\ \cline{2-3} 
                              & logica                & 0.0637          \\ \cline{2-3} 
                              & computadores          & 0.0398          \\ \cline{2-3} 
                              & machine               & 0.0140          \\ \cline{2-3} 
                              & data                  & 0.0498          \\ \cline{2-3} 
                              & matematica            & 0.0634          \\ \cline{2-3} 
                              & processing            & 0.0494          \\ \cline{2-3} 
                              & linear                & 0.0846          \\ \cline{2-3} 
                              & models                & 0.0524          \\ \cline{2-3} 
                              & algebra               & 0.0613          \\ \cline{2-3} 
                              & complexity            & 0.0098          \\ \cline{2-3} 
                              & computational         & 0.0098          \\ \hline
\end{tabular}
\caption{\textit{Tags extracted from books borrowed by the user.}}
\label{tab6}
\end{table}

Table \ref{tab7} presents the lists of area tags containing the terms of books borrowed in period of a specific course, which in this example is Computer Science. The purpose of this list is to identify the course tags that the user is enrolled according to the loans made in each period. 

\begin{table}[]
\centering
\begin{tabular}{|l|l|l|}
\hline
\textbf{Course}                          & \textbf{Period}                & \textbf{Tag - weight} \\ \hline
\multirow{5}{*}{Master Computer Science} & \multirow{5}{*}{Second period} & architecture - 0.1397 \\ \cline{3-3} 
                                         &                                & computer - 0.1397     \\ \cline{3-3} 
                                         &                                & arquitetura - 0.1397  \\ \cline{3-3} 
                                         &                                & computador - 0.1397   \\ \cline{3-3} 
                                         &                                & computadores - 0.1397 \\ \hline
\multirow{5}{*}{Master Computer Science} & \multirow{5}{*}{Third period}  & linear - 0.1590       \\ \cline{3-3} 
                                         &                                & programming - 0.1590  \\ \cline{3-3} 
                                         &                                & programa - 0.1590  \\ \cline{3-3} 
                                         &                                & algebra - 0.1128      \\ \cline{3-3} 
                                         &                                & logica - 0.1128       \\ \hline
\multirow{5}{*}{Master Computer Science} & \multirow{5}{*}{Fourth Period} & analise - 0.1505      \\ \cline{3-3} 
                                         &                                & numeric - 0.1505     \\ \cline{3-3} 
                                         &                                & numerical - 0.1505    \\ \cline{3-3} 
                                         &                                & analysis - 0.1505     \\ \cline{3-3} 
                                         &                                & statistic - 0.1128  \\ \hline
\end{tabular}
\caption{\textit{Tags of the course and their periods.}}
\label{tab7}
\end{table}

With the lists of defined tags, the next step is the grouping of this information to generate recommendations. Thus, the tags are separated into three groups according to their weight: the first group is formed by the tags that have higher than or equal to 80$\%$ of the greater weight, the second group the tags that have weight between 79$\%$ and 50$\%$, and the third group the tags weighing less than or equal to 49$\%$. Using groups consider the following criteria to generate recommendations:
\begin{itemize}
\item If at least two user list group 1 tags are equal the book tags, then the book/document is recommended for the user.
\item If there are not equal tags in the group 1, we look in group 2 if there are at least three tags equal to the book to be recommended. 
\item If in the set of book/document tags there is no tag in group 1 or group 2, then checks whether the group 3. If at least four book tags correspond to the user's tags, then the book/document is recommended.
\end{itemize}

This discretization scheme was used to avoid recommendations that may be irrelevant to the user because a tag that has a lower weight value does not add as much importance as a tag that has a high weight value. Thus, if a recommendation is made considering only a smaller value tag, the probability of irrelevance rating would be higher. Therefore, the tags that have less weight need to be combined to add value on a recommendation.

Then, a list of books and documents is generated as a recommendation for the user to evaluate with a score from 0 (recommendation does not interest) to 3 (recommendation interests). In addition to the recommendations, we calculate the similarity among users, if they have high similarity value (95$\%$), then the books that were borrowed by them will be included in the recommendation list.

Based on feedback, the list of irrelevant tags is formed with the tags of the recommended books/documents, but that the user evaluated with score 0. The user tags list is compared with the list of irrelevant tags, if any irrelevant tag appear on their list will then be removed to avoid irrelevant content recommendations. These steps are repeated in a constant cycle, the more loans and evaluations the user make, the higher the accuracy of the recommendations generated by the system.

\subsubsection{Experiment Evaluation}
To evaluate the recommendations we use precision and recall measures presented in the previous section. The evaluation of the experiment consists of two stages: the first stage we evaluate the recommendations based on the history of the user's loans and in the second stage we use the user feedback to evaluate the recommendations that were suggested. 

We consider relevant books that were borrowed by the user and the system recommended it in the user list. Irrelevant are books that were recommended, but were not borrowed by the user. So, we fill in the confusion matrix to perform the calculations of precision and recall, considering:

\begin{itemize}
\item Nrr = recommended books that were borrowed by the user.
\item Nir = recommended books, but not borrowed by the user.
\item Nrn = books not recommended, but borrowed by the user.
\item Nin = does not apply because there is no way to identify books that are not recommended and were not borrowed by the user, since the recommendation list is generated based on the user's loans in a certain period of time. This value is used if deemed the entire library database, but the increase in processing time would make impracticable the experiments in this study.
\end{itemize}

The experiment 5 implementation that considered 4, 5, and 5 tags, obtained better results with the precision value 0.4490, 0.4352 recall, and F-Score 0.4420, as shown in table \ref{tab8}:

\begin{table}[]
\centering
\begin{tabular}{|l|l|l|l|l|l|}
\hline
\textbf{Experiments} & \multicolumn{5}{l|}{\textbf{Groups}}                                                                                    \\ \hline
Experiment 1         & \multicolumn{5}{l|}{$Group 1 \geq 90\% ; Group 2 \geq 60 \%\ and \leq 89\% ;  Group 3 \leq 59\% $}    \\ \hline
Tags                 & 1, 2 e 3               & 2, 3 e 4               & 3, 4 e 4               & 4, 5 e 5              & 2, 4 e 6             \\ \hline
Precision            & 0.3421                 & 0.3363                 & 0.4081                 & 0.4412                & 0.4271               \\ \hline
Recall               & 0.4375                 & 0.3908                 & 0.4375                 & 0.4372                & 0.4375               \\ \hline
F-Score              & 0.3839                 & 0.3615                 & 0.4223                 & 0.4392                & 0.4322               \\ \hline
\multicolumn{6}{|l|}{}                                                                                                                         \\ \hline
Experiment 2         & \multicolumn{5}{l|}{$Group 1 \geq 85\% ;Group 2 \geq 55\%\ and \leq 84\% ; Group 3:\leq 54\%$}     \\ \hline
Tags                 & 1, 2 e 3               & 2, 3 e 4               & 3, 4 e 4               & 4, 5 e 5              & 2, 4 e 6             \\ \hline
Precision            & 0.3401                 & 0.3362                 & 0.4094                 & 0.4424                & 0.4244               \\ \hline
Recall               & 0.4375                 & 0.3908                 & 0.4375                 & 0.4372                & 0.4375               \\ \hline
F-Score              & 0.3827                 & 0.3615                 & 0.4230                 & 0.4398                & 0.4308               \\ \hline
\multicolumn{6}{|l|}{}                                                                                                                         \\ \hline
Experiment 3         & \multicolumn{5}{l|}{$Group 1  \geq 80\% ; Group 2: \geq 60\%\ and \leq 79\% ; Group 3 \leq 59\%$} \\ \hline
Tags                 & 1, 2 e 3               & 2, 3 e 4               & 3, 4 e 4               & 4, 5 e 5              & 2, 4 e 6             \\ \hline
Precision            & 0.3416                 & 0.3340                 & 0.4077                 & 0.4412                & 0.4228               \\ \hline
Recall               & 0.4355                 & 0.3908                 & 0.4355                 & 0.4352                & 0.4355               \\ \hline
F-Score              & 0.3829                 & 0.3602                 & 0.4211                 & 0.4382                & 0.4291               \\ \hline
\multicolumn{6}{|l|}{}                                                                                                                         \\ \hline
Experiment 4         & \multicolumn{5}{l|}{$Group 1 \geq 80\% ; Group 2 \geq 50\%\ and \leq 79\% ; Group 3 \leq 49\%$} \\ \hline
Tags                 & 1, 2 e 3               & 2, 3 e 4               & 3, 4 e 4               & 4, 5 e 5              & 2, 4 e 6             \\ \hline
Precision            & 0.3469                 & 0.3921                 & 0.3541                 & 0.3825                & 0.3658               \\ \hline
Recall               & 0.4270                 & 0.4270                 & 0.3908                 & 0.3906                & 0.3908               \\ \hline
F-Score              & 0.3828                 & 0.4088                 & 0.3715                 & 0.3865                & 0.3779               \\ \hline
\multicolumn{6}{|l|}{}                                                                                                                         \\ \hline
Experiment 5         & \multicolumn{5}{l|}{$Group 1 \geq 70\% ; Group 2 \geq 40\%\ and \leq 69\% ; Group 3 \leq 39\%$}  \\ \hline
Tags                 & 1, 2 e 3               & 2, 3 e 4               & 3, 4 e 4               & 4, 5 e 5              & 2, 4 e 6             \\ \hline
Precision            & 0.2889                 & 0.3336                 & 0.3564                 & 0.4490                & 0.4120               \\ \hline
Recall               & 0.3908                 & 0.3908                 & 0.3908                 & 0.4352                & 0.4355               \\ \hline
F-Score              & 0.3322                 & 0.3600                 & 0.3729                 & 0.4420                & 0.4234               \\ \hline
\end{tabular}
\caption{\textit{Results of algorithm implementations varying parameters.}}
\label{tab8}
\end{table}

Initially, the threshold measure used considered an exact value of tags. However, many data were disregarded or recovered excessively. So the percentage separation of groups based on the weight of the tags, and the threshold limits were based on \cite{durao2009}, which uses values close to 70$\%$ was applied . A value far above this limit may omit interesting recommendations for the user, while values much below can generate a lot of unnecessary recommendations.

\begin{figure} [!htb]
	\centering
	\includegraphics[width=\textwidth,keepaspectratio]{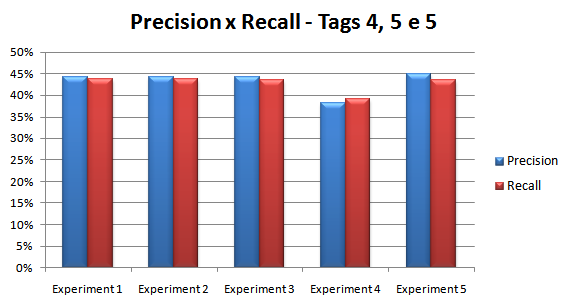}
	\caption{Precision and recall measures for each experiment considering groups with 4, 5 and 5 tags.}
	\label{fig7}
\end{figure}

In figure (\ref{fig7}) are represented the experiments of the algorithm run considering 4, 5, and 5 tags, making possible the analysis of the best cutoff point of the execution that showed better results, in this case obtained by the experiment 5. 

\begin{figure} [!htb]
	\centering
	\includegraphics[width=\textwidth,keepaspectratio]{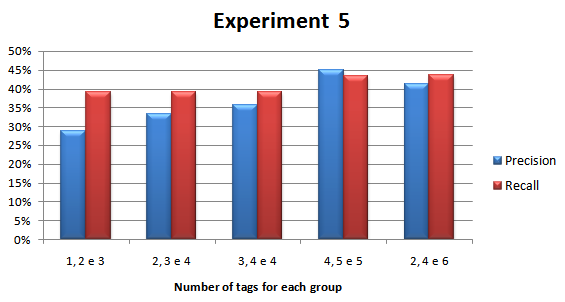}
	\caption{Precision and recall measures for all the five groups of tags in Experiment 5.}
	\label{fig8}
\end{figure}

Analyzing the tag groups in detail in the experiment 5 in figure \ref{fig8}, there is the group 4, 5, and 5 tags obtained better results than others, because some tags when jointly analyzed add a more forceful meaning to the content. A compound term analyzed separately loses the content representation significance. For example, the term ``architecture of computers" form three tags, the term ``of" is removed in preprocessing, and the terms ``architecture" and ``computers" are analyzed separately. If only one term belongs to the list of tags, there is meaning lost because the terms ``architecture" and ``computers" when analyzed separately have varying concepts. 

To improve the accuracy of recommendations, the number of tags used for comparison should increase to a certain limited, as was done in experiment 5 in the group 4, 5, and 6 of figure \ref{fig8}. By increasing the number of tags for comparison, it tends to increase the precision and the recall value may decrease. Still, the recall value maintained a high value in relation to other groups, and the amount of tags 4, 5, and 5 applied to generate the recommendations.

\begin{figure} [!htb]
	\centering
	\includegraphics[width=\textwidth,keepaspectratio]{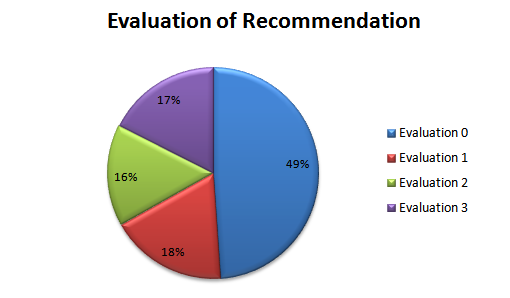}
	\caption{Evaluations made by users on each book document recommended.}
	\label{fig9}
\end{figure}

To find the best threshold the algorithm is run again to generate the list of recommendations (feedback) that will be sent to the user. In this second step, we measure user satisfaction as the recommendations through its feedback. 3,666 evaluations were collected. For the figure \ref{fig9}, 49$\%$ of the assessments made by the users indicate that the content suggestions are not of interest. However, 51$\%$ of the recommended content was assessed at 1, 2, or 3, indicating that these items were suggested well accepted by the user. In this first stage only the history of the user's loan was used to generate the recommendations. The values initially found are not ideal for an effective recommendation system, but are prone to start the first suggestions of content, and can be improved in the next step as the evaluations of users are being collected.

\begin{figure} [!htb]
	\centering
	\includegraphics[width=\textwidth,keepaspectratio]{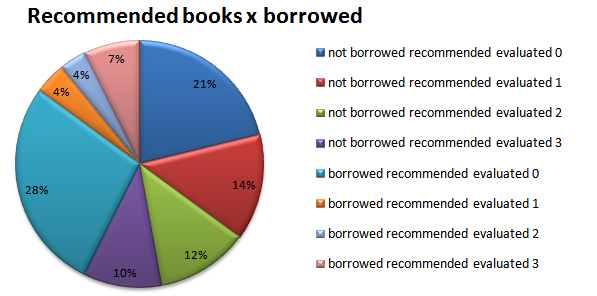}
	\caption{Acceptance of recommendations of books that were or were not borrowed by users.}
	\label{fig10}
\end{figure}

Analyzing data from the first recommendation stage and relating them to the user feedback data, there is the figure \ref{fig10} that 28$\%$ of the books borrowed by the user were recommended, but had negative evaluations. This is because some books that the user make loans are not always of their interest or your preferences change over time. Recommended borrowed books, 15$\%$ (4$\%$, 4$\%$, and 7$\%$) are classified by users as interesting, which demonstrates that the evaluation behavior of users on these items is in accordance with their loan options and were suggested correctly.  

We can also note that 36$\%$ (14$\%$, 12$\%$, and 10$\%$) of books that have not been borrowed had a good acceptance by users. This shows that the algorithm was able to suggest a significant number of books that were unknown to users, but were evaluated positively, showing interest in the recommendations.

\begin{table}[]
\centering
\begin{tabular}{|l|l|l|}
\hline
                    & \textbf{Borrowed} & \textbf{Not borrowed} \\ \hline
\textbf{Relevant}   & 547               & 1326                  \\ \hline
\textbf{Irrelevant} & 1016              & 777                   \\ \hline
\end{tabular}
\caption{\textit{Feedback data.}}
\label{tab9}
\end{table}

Based on table \ref{tab9}, the values calculated for precision, recall, and F-score, which are respectively 34.9$\%$, 29$\%$, and 31$\%$. It appears that the f-score value acquired by user feedback was close to the value of f-score (44$\%$) calculated in the first stage using the historical loans. Thus, we have the recommendations based on the historic representing the user's interest accurately.

\section{Conclusions and Future Work}
In an academic environment the availability and diversity of information is essential to ensure the quality and development of research. Some systems are used for this purpose, though not always provide intelligent technologies to facilitate the search for content in an integrated manner, as is the case of digital repositories and library management systems. Therefore, content recommendation systems are used in academic environment to assist users in the search for new and relevant information. The integration of these services and use of content recommendation techniques on these technologies generate benefits that allow the availability of relevant resources for users and the targeted dissemination of various types of content.

In this sense, we propose a recommendation system that applies a hybrid approach and use tags lists to model the user's profile, adapting to changes in interest over time to allow the indication of new and relevant content. The system extracts from the library database the borrowed books metadata and transforms them into tags that are used to model the user's profile, which receives suggestions for content to be evaluated according to their interest. Items on the recommendation list consist of books from the database library and documents that are retrieved from the institutional repository. This integration provided more content to the user, but in a targeted and controlled manner according to your profile. Through user feedback it was possible to improve the accuracy of the next recommendations. Furthermore, the use of dynamic lists tag allowed the identification of user behavior changes over time. 

The system provided suggestions for items that were not previously explored, but were rated by users as important recommendations. We found that some users who have books of loans have lost interest in the same content. However, through the feedback and lists tags area, we adjust the list of user tags according to changes in interest over time and generate more specific recommendations, adapted to the new user preferences. 

New approaches can be explored in this study to improve the accuracy of the recommendations, such as the use of ontologies and semantics, which improve the quality of representation of the terms in the list of tags. Furthermore, the proposed implementation can be coupled in an integrated search system, which maps the real-time interactions and recovers the user data search using them to add knowledge for future research. Regarding the evaluation of recommendations, other measures may be used on other aspects, such as mean absolute error (MAE) and ROC curve to determine the best classifications and measure the performance of the recommendation system. 



\end{document}